\documentclass[aps,floatfix,onecolumn,a4paper,noshowpacs, nofootinbib,superscriptaddress,11pt]{revtex4}

\usepackage[utf8]{inputenc}
\usepackage{amsmath,amssymb,graphicx,bm}
\usepackage{geometry}

\usepackage{siunitx}
\usepackage{hyperref}
\hypersetup{colorlinks=true, linkcolor=blue, citecolor=blue, urlcolor=blue}
\usepackage{graphicx,float,tikz}
\usepackage[all]{xy}
\usepackage{bm,amsmath,upgreek,bigints}

\usepackage{amssymb}
\usepackage{color}
\usepackage{epsfig,bm}		
\usepackage{graphicx,epstopdf}
\usepackage{subfigure,upgreek}
\usepackage{pdfpages}
\usepackage{multirow}

\newcommand{\beq}{\begin{eqnarray}}
\newcommand{\eeq}{\end{eqnarray}}

\usepackage{bookmark,textgreek}
\usepackage{hyperref,color,xcolor}
\definecolor{lightgray}{gray}{0.95}
\hypersetup{hidelinks,colorlinks=true,breaklinks=true,urlcolor= blue}
\hypersetup{%
    colorlinks = true,
    linkcolor  = blue,
    citecolor = cyan,
  }
\usepackage{siunitx}
\definecolor{headergray}{gray}{0.90}

\usepackage[font=small,labelfont=bf,labelsep=space]{caption}
\usepackage{natbib}
\setcitestyle{square,numbers}

\newcommand\orcidroldao{{\href{https://orcid.org/0000-0003-3978-532X}{\orcidicon}}}

\newcommand{\orcidicon}{%
	\begin{tikzpicture}
	\draw[lime, fill=lime] (0,0)
		circle [radius=0.16]
		node[white] {{\fontfamily{qag}\selectfont \tiny ID}};
	\draw[white, fill=white] (-0.0625,0.095)
		circle [radius=0.007];
	\end{tikzpicture}	\hspace{-2mm}
}

\begin{document}

\title{Hairy black hole solutions in nonlocal quadratic gravity}

\author{Roldao da Rocha\orcidroldao\!\!}
\email{roldao.rocha@ufabc.edu.br}
\affiliation{Federal University of ABC, Center of Mathematics, Santo Andr\'e, S\~ao Paulo, 09580-210, Brazil}

\begin{abstract}
Hairy black hole solutions are constructed within quantum-inspired nonlocal quadratic gravity. Nonlocal effects induce Yukawa screening, which shifts the event horizon inward and modifies both the Bekenstein--Hawking entropy and the Hawking temperature, while also renormalizing the chemical potential. Nonlocal corrections also reduce the magnitude of the negative specific heat, making small hairy black holes more stable, with Helmholtz and Gibbs free energies consistent with the absence of first-order phase transitions.  The nonlocal spin-2 propagator contains, in addition to the massless graviton, a massive pole with positive residue and positive norm. Consequently, hairy black holes in nonlocal quadratic gravity are free of ghost instabilities at the quadratic and classical levels, in the effective field theory.
\end{abstract}

\maketitle

\section{Introduction}
\label{sec1}
The detection of gravitational waves (GWs) by laser-interferometric observatories has advanced the understanding of the strong-field regime of gravity \cite{LIGOScientific:2019fpa}. Emitted during the late stages of binary mergers, GWs enable tests of solutions to the Einstein field equations and their extensions. In particular, the ringdown phase of compact binary coalescences is governed by quasinormal modes (QNMs), which are sensitive to deviations from the Kerr and Reissner--Nordström geometries. Consequently, GW observations provide a powerful probe of black hole hair and short-range modifications of gravity.  

According to the classical no-hair theorems, stationary, asymptotically flat
black holes in four-dimensional Einstein--Maxwell theory are completely
characterized by their mass, electric charge, and angular momentum.
All additional information about the matter content that formed the black hole is hidden behind the event horizon, leading to the widely held belief that black holes admit no independent hair \cite{Israel:1967za}.
The assumptions underlying no-hair theorems can be evaded in a wide class of extended gravitational theories. In particular, the presence of additional scalar degrees of freedom or higher-derivative interactions allows for black hole
solutions endowed with nontrivial external field configurations, commonly
referred to as hairy black holes \cite{Bekenstein:1971hc}.
Prominent developments include hairy black holes in gravitational decoupling (GD) \cite{Ovalle:2020kpd,Ovalle:2017fgl,Ovalle:2018gic,Rincon:2019jal,Avalos:2023ywb,Albalahi:2024vpy,Sharif:2025oeu,Andrade:2025xxl,Berkimbayev:2025ail}, which have also been applied to approach  the dark matter problem \cite{Maurya:2024zao,Almatroud:2025xqq,Maurya:2025kto}. GD hairy black holes carry additional degrees of freedom not associated with quasilocal conserved quantities. Analyzing primary hair provides key input for the consistent analytical construction of stellar configurations and black holes in GR and quantum gravity approaches \cite{Casadio:2017sze}. GD hairy black holes were also constructed in AdS spacetimes, as mentioned in Refs.  \cite{Zhang:2022niv,Lin:2024ubg}.  Observationally, the QNM emission and related signatures of hairy black holes have been studied in Refs.~\cite{Guimaraes:2025jsh,Cavalcanti:2022cga,Zhidenko:2005mv,Ribeiro:2025ohn,Yang:2022ifo,Avalos:2023jeh,Al-Badawi:2024iax,Tello-Ortiz:2024mqg}.   Hairy black holes have been extensively investigated through gravitational lensing analyses \cite{Cavalcanti:2016mbe,Wang:2025hzu,Liang:2024xif,Islam:2021dyk}, redshift measurements \cite{Gabbanelli:2018bhs}, and studies within analog gravity  \cite{Casadio:2024uwj,Li:2022hkq}. 
The thermal stability of hairy black holes was addressed in Refs. \cite{Ditta:2023arv,Mansour:2024mdg,Mahapatra:2022xea,Misyura:2024fho}, with astrophysical accretion mechanisms proposed in Ref. \cite{Rehman:2023eor}.
Hairy black hole solutions were also studied in the infrared (IR) regime of AdS/CFT \cite{daRocha:2017cxu,Meert:2020sqv,Takahashi:2019oxz,Ovalle:2013vna,daRocha:2012pt}, including holographic entanglement entropy approaches \cite{daRocha:2020gee}.  Other relevant examples include rotating black holes with scalar hair in
Einstein--Klein--Gordon and Einstein--Maxwell-dilaton theories  \cite{Herdeiro:2014goa,Khalil:2018aaj,Kleihaus:2015iea} and scalar hairy black holes \cite{Doneva:2017bvd,Ballesteros:2023muf,Ganchev:2017uuo,Tahamtan:2020mbb,Bravo-Gaete:2020lzs}. Additionally, hairy black holes have been investigated in the context of Lovelock gravity \cite{Estrada:2019aeh,Hennigar:2016ekz}, Brans--Dicke gravity \cite{Sharif:2020lbt}, Riemann--Cartan spacetimes \cite{Priyadarshinee:2021rch,Naseer:2025fwm}, extensions of Einstein--Maxwell gravity \cite{Mahapatra:2020wym,Pradhan:2025dno}, and in cubic gravity \cite{Dykaar:2017mba}.

Exploring gravity beyond GR remains a central pursuit in modern theoretical physics, aimed also at deepening our understanding of the fundamental structure of spacetime. Despite its success, GR requires extensions in high-energy or high-curvature regimes, where quantum effects become significant \cite{Pasqua:2014owa}. A relevant extension of GR comprises nonlocal gravity, which has curvature invariants that interact via nonlocal operators \cite{Calmet:2024pve}, yielding ultraviolet (UV) softening implemented by UV regulators in the propagator and regulating ghost-like degrees of freedom \cite{Kuntz:2019qcf,Shapiro:2015uxa,Casadio:2021rwj}. 
 Nonlocal approaches to gravity are characterized by the appearance of nonlocal operators, typically involving d'Alembertian kernels,
which modify the gravitational dynamics over a finite range of scales
\cite{Maggiore:2014sia}.
Quantum extensions of the Reissner-Nordstr{\"o}m solution were proposed in Refs. \cite{Antonelli:2025mcv,Casadio:2022ndh}.  
Black hole solutions in nonlocal gravity have begun to attract increasing
interest \cite{dePaulaNetto:2023cjw,Dimitrijevic:2022fhj,Kolar:2021rfl,Modesto:2017sdr}.
Perturbative and exact solutions have been constructed in  nonlocal gravity, revealing modifications to the horizon structure, effective matter sources, and asymptotic behavior
\cite{Frolov:2015bia,Buoninfante:2018xiw,Edholm:2016hbt,Fernandes:2017vvo}.
These studies indicate that nonlocal effects can mimic extended matter
distributions and generate finite-range corrections to the metric, raising the
possibility that nonlocal gravity naturally supports black hole hair \cite{Maggiore:2016fbn}.

This work focuses on constructing hairy black hole solutions with Yukawa-like terms that decay at a scale set by the effective mass parameter $\upmu$, which appears in the nonlocal operator, arising from a nonlocal quadratic extension of Einstein--Maxwell gravity.  The hairy black hole thermodynamic framework is developed and analyzed in detail. The nonlocal propagator exhibits, besides a massless graviton, an additional massive spin-2 pole with positive residue and positive norm, ensuring the absence of ghost instabilities at the quadratic level. Consequently, the resulting hairy black hole metrics remain well defined within the effective field theory (EFT) regime.  

This paper is organized as follows.  Section \ref{sec3} introduces the quantum-inspired nonlocal quadratic gravity action, whose Einstein field equations are solved perturbatively.  Nonlocal effects induce  Yukawa screening, shrink the event horizon, and modify the Hawking temperature and Bekenstein--Hawking entropy. They also decrease the magnitude of the specific heat, renormalize the chemical potential, and introduce nonlocal corrections to the Helmholtz and Gibbs free energies. Section \ref{sec5} shows that the nonlocal propagator contains, besides a massless graviton, an additional massive spin-2 pole with positive norm and positive residue, ensuring that the full metric remains well defined in the EFT. Section \ref{sec6} presents concluding remarks.

\section{Black hole solutions in nonlocal quadratic gravity}
\label{sec3}

Hairy black hole solutions will be derived in a nonlocal quadratic gravitational theory with action\footnote{Hereon reduced Planck units will be adopted, setting $8 \pi G = 1$.}
\begin{equation}\label{act}
\!\!\!\!S =  \int d^{4}x \sqrt{-g} \left[
\frac{1}{2}R + \alpha \left(
R_{\mu\nu} (\Box - \upmu^{2})^{-1} R^{\mu\nu}
- \frac{1}{4} R (\Box - \upmu^{2})^{-1} R
\right)
- \frac{1}{4} F_{\mu\nu} F^{\mu\nu}\right],
\end{equation}
where $\alpha$ is the parameter governing nonlocal gravity effects, $\upmu$ is the mass scale setting the distance beyond which nonlocal effects are suppressed, and $F_{\mu\nu}$ is the electromagnetic field strength.  
The nonlocal terms in (\ref{act}) modify Einstein--Maxwell gravity by incorporating quantum-inspired corrections, which arise in the low-energy effective field theory.  The specific combination of Ricci-tensor and scalar curvature terms in the action (\ref{act}) ensures a controlled deformation of GR that precludes a propagating ghost at mass $\sim\upmu$  \cite{Barvinsky:2003kg,DeFelice:2014kma,Shapiro:2008sf}.  

The field equations from the nonlocal quadratic gravity action (\ref{act}) are given by 
\begin{equation}\label{gtt1}
R_{\mu\nu}-\frac12R g_{\mu\nu} + \alpha H_{\mu\nu}^{\scalebox{0.57}{$\textsc{NL}$}} =  T_{\mu\nu}^{{\scalebox{0.5}{$\textsc{(EM)}$}}},
\end{equation}
where 
\beq\label{emm1}
\displaystyle T_{\mu\nu}^{\scalebox{0.5}{$\textsc{(EM)}$}}=F_{\mu\sigma} F_\nu{}^\sigma - \frac{1}{4} g_{\mu\nu} F_{\sigma\beta} F^{\sigma\beta}
\eeq is the electromagnetic energy-momentum tensor, and the nonlocal (NL) contribution in Eq. (\ref{gtt1}) reads
\begin{equation}\label{hnl}
H^{\scalebox{0.57}{$\textsc{NL}$}}_{\mu\nu}
=-\frac{2}{\sqrt{-g}}\frac{\delta S_{\scalebox{0.57}{$\textsc{NL}$}}}{\delta g^{\mu\nu}},
\end{equation}
 taking into account the nonlocal part of the action (\ref{act}): 
\begin{equation}\label{nl1}
S_{\scalebox{0.57}{$\textsc{NL}$}}
=\alpha\int d^4x\,\sqrt{-g}
\left[
R_{\mu\nu}(\Box-\upmu^{2})^{-1} R^{\mu\nu}
-\frac14 R(\Box-\upmu^{2})^{-1} R
\right].
\end{equation}
One can split Eq. (\ref{hnl}) as $H^{\scalebox{0.57}{$\textsc{NL}$}}_{\mu\nu}
=
H^{(1)}_{\mu\nu}
-\displaystyle\frac14
H^{(2)}_{\mu\nu}$, where  the Ricci-tensor part of the nonlocal action reads\footnote{In the last line of Eq. (\ref{sh1}), the notation for the antisymmetrization of the indexes $\mu$ and $\nu$ is used in the second term.}
\beq\label{sh1}
H_{\mu\nu}^{(1)}
\!&\!\equiv\!&\! -\frac{2}{\sqrt{-g}} \frac{\delta}{\delta g^{\mu\nu}}
\left(\int d^4 x \, \sqrt{-g} \, 
R_{\rho\sigma} (\Box - \upmu^{2})^{-1} R^{\rho\sigma} \right)\nonumber\\&\!=\!&\! -4 R_{\mu\rho} (\Box-\upmu^2)^{-1} R_\nu{}^\rho 
- \frac12 g_{\mu\nu} R_{\rho\sigma} (\Box-\upmu^2)^{-1} R^{\rho\sigma} 
+ \nabla_\mu \nabla_\nu \big[(\Box-\upmu^2)^{-1} R\big] \nonumber\\
&& - \Box \big[(\Box-\upmu^2)^{-1} R_{\mu\nu}\big] 
\!-\! 2 \nabla_\rho R_{\mu\sigma}  \nabla^\rho \big[(\Box-\upmu^2)^{-1} R_\nu{}^\sigma\big] 
\!-\! 2 \nabla_\rho R_{\nu\sigma}  \nabla^\rho \big[(\Box\!-\!\upmu^2)^{-1} R_\mu{}^\sigma\big] \nonumber\\
&& + 2 R_{\mu\rho\nu\sigma} (\Box-\upmu^2)^{-1} R^{\rho\sigma} 
- 2(\Box-\upmu^2)^{-1} \nabla_\rho R_{(\mu|\sigma} \, \nabla^\rho (\Box-\upmu^2)^{-1} R_{|\nu)}{}^\sigma,\eeq
whereas the scalar curvature part of the nonlocal action is given by 
\beq
\label{sh2}
H_{\mu\nu}^{(2)} \!&\!\equiv\!&\! -\frac{2}{\sqrt{-g}} \frac{\delta}{\delta g^{\mu\nu}}
\left(\int d^4 x \, \sqrt{-g} \, 
R(\Box - \upmu^{2})^{-1} R \right)\nonumber\\&\!=\!&
- \frac12 R_{\mu\nu} (\Box-\upmu^2)^{-1} R
+ \nabla_\mu \nabla_\nu (\Box-\upmu^2)^{-1} R
- g_{\mu\nu} \Box (\Box-\upmu^2)^{-1} R \nonumber \\
&&
+ \frac12 (\Box-\upmu^2)^{-1}
\!\left[ \nabla_{(\mu} R \, \nabla_{\nu)} (\Box-\upmu^2)^{-1} R \right]
- \frac14 g_{\mu\nu} (\Box-\upmu^2)^{-1}
\!\left[ \nabla_\rho R \nabla^\rho (\Box-\upmu^2)^{-1} R \right]\nonumber \\
&&+ \frac14 g_{\mu\nu} R (\Box-\upmu^2)^{-1} R
- \frac12 (\Box-\upmu^2)^{-1}
\!\left[ R \nabla_\mu \nabla_\nu (\Box-\upmu^2)^{-1} R \right].\eeq 
\noindent The nonlocal structures (\ref{sh1}, \ref{sh2}) naturally arise in the one-loop effective action of gravity \cite{Barvinsky:1994cg,Buccio:2024hys,Barvinsky:2003kg}.
Integrating out massive fields or gravitons generates the nonlocal quadratic operators in the action (\ref{act}), 
with a loop-suppressed coefficient $\alpha \sim (16\pi^{2})^{-1}$ encoding the strength of the quantum corrections, whereas the scale $\upmu$ corresponds to the mass of the
integrated-out field. The nonlocal quadratic gravity action (\ref{act}) describes a low-energy EFT in a quantum gravity approach and can also define a consistent classical theory. 

In Subsecs. \ref{lo} and \ref{sec4} we perform a perturbative expansion in $\alpha$, treating the nonlocal corrections iteratively via the  Yukawa Green function. The nonlocal quadratic gravity effects on the thermodynamics of hairy black holes will be addressed in Subsec. \ref{nlo}. 

\subsection{Leading-order nonlocal metric corrections}
\label{lo}
The nonlocal operator $(\Box - \upmu^2)^{-1}$ prevents a closed-form analytic solution of the equations of motion derived from the action (\ref{act}). Therefore, we implement a perturbative expansion in the small parameter $\alpha$, which encodes the strength of the nonlocal corrections relative to Einstein--Maxwell gravity. 

For a static, spherically symmetric metric 
\beq\label{sss}
ds^2 = f(r) dt^2 - f(r)^{-1} dr^2 - r^2 d\Omega^2,
\eeq
the Maxwell equations, with electromagnetic potential ansatz
$A_\mu  = A_t(r)\delta_{\mu}^0$, read 
\begin{equation}
F_{tr} = \frac{Q}{r^{2}}, \quad\qquad F_{\theta\phi} = 0 ,
\end{equation}
where $Q$ is the conserved electric charge.

The metric function $f(r)$ in (\ref{sss}) 
can be obtained by solving the $tt$-component of the nonlocal field equations (\ref{gtt1}). For it, we need the term 
\begin{equation}
T^{t}{}_{t}{}^{\scalebox{0.5}{$\textsc{(EM)}$}}
=
-\frac{Q^{2}}{2 r^{4}}
\end{equation}
from Eq. (\ref{emm1}). Therefore, the $tt$-component of (\ref{gtt1}) reduces to the following ODE: 
\begin{equation}\label{itt}
\frac{d}{dr} \left(r f(r) \right) = 1 + \frac{Q^2}{r^2} + \alpha r^2  \Upgamma_{\scalebox{0.57}{$\textsc{NL}$}}(r),
\end{equation}
where the term 
\beq
\Upgamma_{\scalebox{0.57}{$\textsc{NL}$}}(r)
=
M \upmu^{3} e^{-\upmu r}
+ \frac{M\upmu^{2}}{2r} e^{-3 \upmu r}
\eeq
 encodes the nonlocal contributions. Integrating
term by term in Eq. (\ref{itt}) yields 
\begin{eqnarray}
r f(r) &=&r - \frac{Q^2}{r} +  {I_{\scalebox{0.57}{$\textsc{NL}$}}} + C,\label{rfr}
\end{eqnarray}
for $C$ denoting an integration constant, 
where the nonlocal integral yields 
\begin{eqnarray}
\!\!\!\!\!\!\!\!\!\!\!\!
I_{\scalebox{0.57}{$\textsc{NL}$}}
&=&
\alpha \int_0^r d\mathsf r\, \mathsf r^{2}\,
\Upgamma_{\scalebox{0.57}{$\textsc{NL}$}}(\mathsf r)
\nonumber\\[1ex]
&=&
-\alpha M \upmu \, e^{-\upmu r}
\left(
r^{2} + \frac{2 r}{\upmu} + \frac{2}{\upmu^{2}}
\right)
- \frac{\alpha M}{6 \upmu}
e^{-3 \upmu r}
\left(
r + \frac{1}{3 \upmu}
\right),
\label{subl}
\end{eqnarray}
The integral (\ref{subl}) leads to the most general metric solution at leading order:
\begin{equation}\label{ms1}
f(r) = 1 - \frac{2 M}{r} + \frac{Q^2}{r^2} 
+ {\alpha M\upmu} e^{-\upmu r} \left( r + \frac{2}{\upmu} + \frac{2}{\upmu^2 r} \right)
- \frac{\alpha M}{18\upmu r} e^{-3 \upmu r} \left( 3 \upmu r + 1 \right).
\end{equation}
The black hole hair corresponds to the nontrivial modifications of the metric (\ref{ms1}), beyond the Reissner--Nordstr\"om terms, induced by the nonlocal quadratic curvature terms, which cannot be absorbed into $M$ or $Q$.

\subsection{Next-to-leading order nonlocal corrections to the Reissner--Nordström metric}
\label{sec4}
We now compute the next-to-leading order ($\mathcal{O}(\alpha^2)$)  corrections to the metric ansatz (\ref{sss}), building on the zeroth- and first-order solutions:
\beq\label{fullm}
f(r) = f_0(r) + \alpha f_1(r) + \alpha^2 f_2(r),
\eeq
where 
\beq\label{1s}
f_0(r) =1 - \frac{2 M}{r} + \frac{Q^2}{r^2}
\eeq corresponds to the Reissner--Nordström term,  while for the leading-order metric coefficient (\ref{ms1}), we denote  
\beq\label{2s}
 \!\!\!\!\!\!\!\!\!\!\!\!\!\!\!f_1(r)=\displaystyle{M \upmu} e^{-\upmu r} \left( r + \frac{2}{\upmu} + \frac{2}{\upmu^2 r} \right)
- \frac{M}{18\upmu r} e^{-3 \upmu r} \left( 3 \upmu r + 1 \right),\label{fgm2}
\eeq

To implement the next-to-leading order term $f_2(r)$, the convolution involving the scalar curvature 
admits the solution 
\beq\label{yr}
Y(r) &\equiv& (\Box - \upmu^2)^{-1} R(r)= \int_0^\infty ds \, s^2 \, G(r,s) \, R(s),
\eeq
where the Yukawa-type Green function 
\begin{equation}\label{gf}
G(r,s) = \frac{e^{-\upmu|r-s|}}{4\pi r s}, 
\end{equation} 
satisfies
\begin{equation}
(\Box - \upmu^{2}) G(r,s) = \frac{\delta(r-s)}{r^2}.
\end{equation}
Using the candidate metric (\ref{ms1}), the scalar curvature reads 
\beq
\!\!\!\!\!\!\!\!\!\!\!\!R(r)&\!=\!& - f''(r) \!-\! \frac{4}{r} f'(r) \!+\! \frac{2}{r^2} \left( 1 - f(r) \right)= \alpha M \frac{e^{-\upmu r}}{r}
\left(\upmu - \frac{3}{r}\right)+\alpha  \frac{M}{2} e^{-3 \upmu r} \left(3\upmu - \frac{3}{r} \right).\label{rr11}
\eeq
The last term in Eq. (\ref{rr11}) is small compared to the $\mathcal{O}\left(e^{-\upmu r}\right)$ term even when $\upmu r\sim 1$, being suppressed when $\upmu r\gg1$. 
Hence, the convolution integral (\ref{yr}) yields
\beq
Y(r)=\!\alpha\!\left[
\frac{1}{r}\!\left(\frac{3M}{2}e^{-\upmu r}\!\!\left(\frac{e^{-2\upmu r}}4\!-\!\frac12
\!+\!\mathrm{Ei}(-2\upmu r)\!-\!e^{2\upmu r}\mathrm{Ei}(-4\upmu r) \right)\!\!\right)
\!+\!\frac{2}{\upmu r}\!+\!1\!+\!e^{-\upmu r}\mathrm{Ei}(-\upmu r)
\right],
\eeq
where $\mathrm{Ei}(a) = {\small\displaystyle- \int_{a}^{\infty} dt\,\frac{e^{-t}}{t}},$ denotes the exponential integral. 

Similarly, for the radial Ricci tensor component
\beq
\label{rrr2}
R_{rr}(r)
=
-\frac{f''(r)}{2 f(r)}
-\frac{f'(r)}{r f(r)}
+\frac{f'(r)^2}{4 f(r)^2},
\eeq
the convolution reads 
\beq\label{con}
Y_{rr}(r)
&\equiv& (\Box - \upmu^2)^{-1} R_{rr}(r)
= \int_0^\infty ds \, s^2 \, G(r,s) \, R_{rr}(s).
\eeq
To obtain the next-to-leading correction to the metric, the radial component of the field equations (\ref{gtt1}), including nonlocal corrections, can be written in the form
\begin{equation}\label{radial-eq}
r f'(r) + f(r) - 1 = r^2  T_{rr}^{{\scalebox{0.57}{$\textsc{eff}$}}}(r)
\end{equation}
where  
\begin{equation}\label{k2}
T_{rr}^{{\scalebox{0.57}{$\textsc{eff}$}}}(r)\equiv K_2(r) = -2 R_r{}^r(r) Y_{rr}(r) + \frac{1}{2} R_{rr}(r)  Y(r) + \frac{e^{-2\upmu r}}{4r^{4}}
\left(
\upmu^{2} r^{2} + 2\upmu r + 2
\right)
\end{equation}
represents the effective radial source. 

From Eq. \eqref{radial-eq}, the second-order correction $f_2(r)$ in Eq. (\ref{fullm}) satisfies the ODE $\big(r f_2(r) \big)^\prime = K_2(r),$ whose integration yields 
\beq\label{ri}
f_2(r)
= \frac{1}{r} \int_0^r  d\mathsf{r}\,K_2(\mathsf{r}).
\eeq
One substitutes Eqs. (\ref{yr}, \ref{rr11}, \ref{rrr2}, \ref{con}) into Eq. (\ref{k2}),  performs the radial integral (\ref{ri}), and requires asymptotically flatness and that $\displaystyle\lim_{r \to \infty}f_2(r) = 0$, yielding the full metric \eqref{fullm} to be given by 
\beq
\label{fm1}
\!\!\!\!\!\!\!\!\!\!\!\!\!\!\!\!\!\!\!f(r) \!&\!=\!&\! 1 - \frac{2 M}{r} + \frac{Q^2}{r^2} 
+ {\alpha M \upmu} e^{-\upmu r} \left( r + \frac{2}{\upmu} + \frac{2}{\upmu^2 r} \right)
- \frac{\alpha M}{18\upmu r} e^{-3 \upmu r} \left( 3 \upmu r + 1 \right).
 \nonumber\\
&&+ \frac{\alpha^2}{r^2} \Bigg[\!
\left( \frac{5 M^2}{2} \!+\! \frac{M Q^2}{r} \!+\! \frac{Q^4}{r^2} \right) e^{-2 \upmu r} 
\!+\! \left({M^2}  \mathrm{Ei}(-\upmu r) \!+\! \frac{M Q^2}{r} e^{-\upmu r} \right) e^{-\upmu r} 
\!+\! \frac{3M^2}{5} e^{-3 \upmu r} 
\Bigg].\eeq
\begin{figure}[H]
    \centering
    \includegraphics[width=8cm]{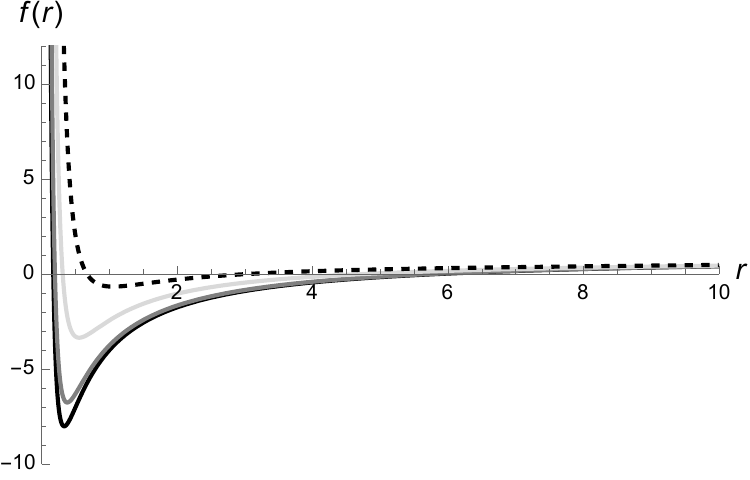}
    \caption{\raggedright Metric function (\ref{fm1}) as a function of the radial coordinate, for $M=3 M_\odot$, $Q= M/3$, and $\upmu=0.4$. The black line regards the Reissner--Nordstr\"om case ($\alpha = 0$), the gray line refers to $\alpha = 0.01$; the light-gray line is plotted for $\alpha = 0.1$, and the dashed line shows the $\alpha = 0.2$ case.}
    \label{fig0}
\end{figure}

\begin{figure}[H]
    \centering
    \includegraphics[width=8cm]{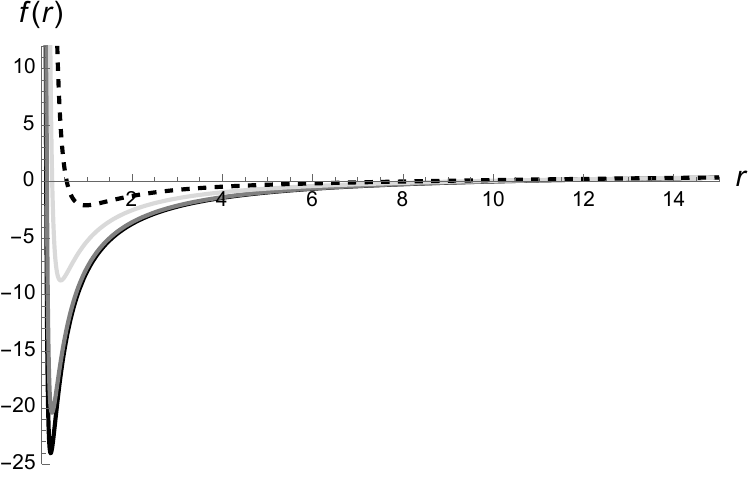}
    \caption{\raggedright Metric function (\ref{fm1}) as a function of the radial coordinate, for $M=10^2 M_\odot$, $Q= M/3$, and $\upmu=0.4$. The black line regards the Reissner--Nordstr\"om case ($\alpha = 0$), the gray line refers to $\alpha = 0.01$; the light-gray line is plotted for $\alpha = 0.1$, and the dashed line shows the $\alpha = 0.2$ case.}
    \label{fig00}
\end{figure}  
\noindent Figs. \ref{fig0} and \ref{fig00} show that the hairy black hole metric (\ref{fm1}) has two horizons, similar to the two-horizon structure $r_\pm = M \pm \sqrt{M^2 - Q^2}$  in the Reissner--Nordstr\"om case. However, the inner Cauchy horizon is shifted outward as $\alpha$ increases, compared to the Reissner--Nordstr\"om case. In contrast, the event horizon is shifted inward as $\alpha$ increases.

\subsection{Thermodynamics of hairy black holes in nonlocal quadratic gravity}
\label{nlo}
In the context of the nonlocal quadratic gravity action (\ref{act}), the classical Reissner--Nordstr\"om event horizon 
\begin{equation}
r_+ = M + \sqrt{M^2 - Q^2} 
\simeq 2 M \left( 1 - \frac{Q^2}{4 M^2} - \frac{Q^4}{16 M^4} \right) 
+ \mathcal{O}\!\left( \frac{Q^6}{M^6} \right),
\end{equation}
is altered to include nonlocal corrections from the metric coefficient (\ref{fm1}). From now on we consider terms up to $\mathcal{O}\left(\frac{Q^6}{M^6}, \alpha^3 \right)$. The event horizon of the hairy black holes metric (\ref{fm1}) reads    
\begin{equation}\label{rh1}
\!\!r_\textsc{h} = 2 M \Bigg[ 1 \!-\! \frac{Q^2}{8M^2} \!-\!  \frac{Q^4}{64M^4} \!-\! \frac{\alpha}{2} e^{-2 \upmu M}\! \left(\!2 \upmu M \!+\! 1 \!+\! \frac{1}{2 \upmu M}\! \right) \!-\! \frac{\alpha^2}{2} \mathrm{Ei}(-2 \upmu M)\! \left(\!1 \!+\! \frac{Q^2}{4 M^2} \right)\! \Bigg].
\end{equation}
As pointed out in Figs. \ref{fig0} and \ref{fig00}, the nonlocal corrections shift the event horizon inward, reflecting the short-range smearing effect of nonlocality, which effectively redistributes the black hole mass near the horizon. This behavior can be understood as a Yukawa screening effect. 

\begin{figure}[H]
    \centering
    \includegraphics[width=8cm]{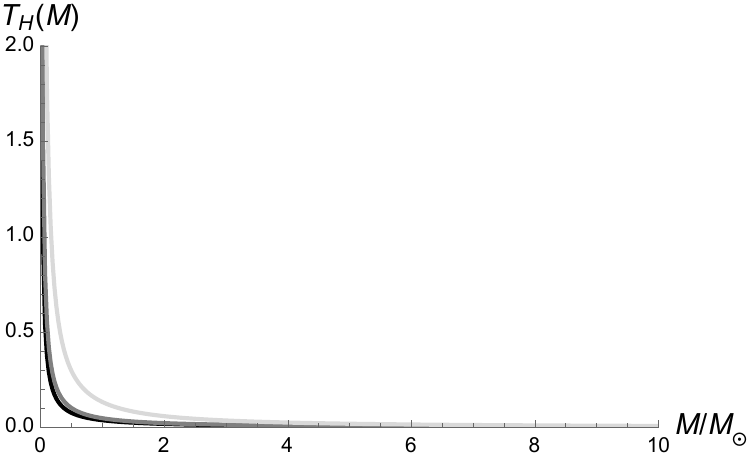}
    \caption{\raggedright Hawking temperature ($\times\, 10^{-7}$ K), as a function of $M/M_\odot$, for $Q= M/3$, and $\upmu=0.001$. The black line regards the Reissner--Nordstr\"om case ($\alpha = 0$), the gray line refers to $\alpha = 0.2$; the light-gray line is plotted for $\alpha = 0.5$.}
    \label{fig2}
\end{figure}

The Hawking temperature reads
\beq\label{th}
T_H &=& \frac{f'(r_\textsc{h})}{4 \pi}=\frac{1}{8 \pi M} \Bigg[ 1 -  \frac{Q^2}{4M^2} -  \frac{Q^4}{16 M^4} \Bigg]- \frac{\alpha}{4 \pi} \Bigg[ \upmu e^{-2 \upmu M} \left( 2 \upmu M + 1 + \frac{1}{2 \upmu M} \right) \Bigg] \nonumber\\
&&\qquad\qquad\qquad\qquad\qquad\qquad\qquad\qquad\displaystyle- \frac{\alpha^2}{4 \pi} \Bigg[ \frac{1}{M}\mathrm{Ei}(-2 \upmu M) \left( 1 + \frac{Q^2}{4 M^2} \right) \Bigg].
\eeq
The plot of the Hawking temperature as a function of the black hole mass is displayed in Fig. \ref{fig2}.
 The value $\upmu=0.01$ corresponds to interaction ranges $\sim 10^2$  Schwarzschild radii for a solar-mass hairy black hole. This ensures nonlocal corrections remain perturbative and affect the horizon-influenced region, where spacetime curvature and redshift are still dominated by the hairy black hole and not by asymptotic flatness.
\noindent One can notice in Fig. \ref{fig2} that the Hawking temperature increases as $\alpha$ increases, for each fixed value of $M$. This implies that the black hole emits more thermal radiation for larger values of $\alpha$, which modifies the horizon properties, making the black hole effectively hotter even without changing its mass.

For the nonlocally-corrected metric (\ref{fm1}), corresponding to the event horizon (\ref{rh1}), the Bekenstein--Hawking entropy  reads 
\beq\label{bh2}
\!\!\!\!\!\!\!S \!=\! 32 \pi^2 M^2 \Bigg[1\!-\!  \frac{Q^2}{2M^2} \!+\!  \frac{Q^4}{16M^4} 
\!-\! \alpha e^{-2 \upmu M} \left( 2 \upmu M \!+\! 1 \!+\! \frac{1}{2 \upmu M} \right)- \alpha^2  \mathrm{Ei}(-2 \upmu M) \left(1 \!+\! \frac{Q^2}{4 M^2} \right) \Bigg].
\eeq
The leading terms reproduce the standard Reissner--Nordstr\"om entropy, while the hairy contributions encode nonlocal corrections. The linear
$\alpha$ term directly reduces the area-law contribution, while the
$\alpha^{2}$ correction is exponentially suppressed for $\upmu M\gg1$ and remains subleading at large $M$.  For $\upmu M=\mathcal O(1)$, the exponential damping is weakened, but the quadratic term is still controlled by an additional power of $\alpha$ and does
not compete with the leading correction in the perturbative regime.  

Now, the specific heat of the black hole solution (\ref{fm1}) can also be obtained, as 
\beq\label{cv1}
C_V^{-1} &=& 
-\frac{1}{8\pi M^2} 
+ \frac{3 Q^2}{32 \pi M^4} 
+ \frac{5 Q^4}{128 \pi M^6} 
+ \frac{\alpha}{4 \pi} \upmu e^{-2\upmu M} \left( 4 \upmu^2 M + \frac{1}{M} + \frac{1}{2 \upmu M^2} \right) \nonumber\\&&
\qquad\qquad+ \frac{\alpha^2}{4 \pi} \left[ \frac{\mathrm{Ei}(-2 \upmu M)}{M^2} \left( 1 + \frac{3 Q^2}{4 M^2} \right) + \frac{2 \upmu e^{-2 \upmu M}}{M} \left( 1 + \frac{Q^2}{4 M^2}\right) \right].
\eeq Positive corrections from $\alpha$ and $\alpha^2$ terms in (\ref{cv1}) 
 reduce the magnitude of the negative specific heat of a Reissner--Nordstr\"om black hole of the same mass and charge, slightly stabilizing small hairy black holes due to nonlocal effects.

Using the Hawking temperature \eqref{th} and the Bekenstein--Hawking entropy \eqref{bh2}, the Helmholtz free energy can now be calculated as
\beq
\!\!\!\!\!\!\!\!\!\!\!\!\!\!\!\!\!\mathcal{F} \!&\!=\!&\! M - T_H S \nonumber\\
\!\!\!\!\!\!\!\!\!\!\!&\!=\!&\!
\frac{M}{2}
\!+\! \frac{3 Q^2}{8 M}
\!-\! \frac{5 Q^4}{64 M^3}\!+\! \frac{\alpha M}{2}
\left[
e^{-2 \upmu M}
\left(
2 \upmu M \!+\! 1 \!+\! \frac{1}{2 \upmu M}
\right)
\right]
\!+\! \frac{\alpha^2 M}{2}
\mathrm{Ei}(-2 \upmu M)
\left(
1 \!+\! \frac{Q^2}{4 M^2}
\right).
\label{helmholtz}
\eeq
The leading contribution in Eq. (\ref{helmholtz}) reproduces the Helmholtz free energy of the Reissner--Nordstr\"om solution. 
For $\upmu M\gg 1$, both the linear and quadratic terms are exponentially suppressed. Still, powers of $M$ enhance the $\alpha$ contribution compared to the $\alpha^{2}$ term, which is further suppressed and
negative, only providing a small compensating correction to the leading nonlocal effect.  When $\upmu M=\mathcal O(1)$, the exponential suppression is mild, yet the linear correction dominates as long as the perturbative regime applies.

The chemical potential $\Upphi$ at the event horizon, using Eq. \eqref{bh2}, can be written as 
\beq
\Upphi & \equiv& \left( \frac{\partial M}{\partial Q} \right)_{\!S}=
\frac{Q}{2M}
\left(
1 + \frac{Q^2}{4M^2}
\right)
\left[
1
- \alpha^2 \mathrm{Ei}(-2\upmu M)
\right],
\label{phi}
\eeq
where the leading-order $\alpha$ contribution vanishes.

With Eqs.~\eqref{th}, \eqref{bh2}, and \eqref{phi}, the first law of black hole thermodynamics 
\beq
dM = T_H dS + \Upphi dQ\eeq  is satisfied, providing a nontrivial consistency check of the hairy black hole thermodynamics in the context of nonlocal quadratic gravity. 

In the grand canonical ensemble, the Gibbs free energy reads 
\beq
\mathcal{G} &=& M - T_H S - \Upphi Q =
\frac{M}{2}
- \frac{Q^2}{8 M}
- \frac{5 Q^4}{64 M^3}
+ \frac{\alpha M}{2}
\left[
e^{-2 \upmu M}
\left(
2 \upmu M + 1 + \frac{1}{2 \upmu M}
\right)
\right]
\nonumber\\
&&\quad\qquad\qquad\qquad\qquad\qquad\qquad\qquad\qquad+ \frac{\alpha^2 M}{2}
\mathrm{Ei}(-2 \upmu M)
\left(
1 - \frac{Q^2}{4 M^2}
\right).
\label{gibbs}
\eeq
Besides, the quantity $\displaystyle{\partial^2 \mathcal{G}}/{\partial T_H^2} 
= - {C_V}/{T_H}$ is finite and analytical, supporting the absence of first-order phase transitions, with a smooth crossover between classical Einstein--Maxwell gravity and nonlocal regimes.

The thermodynamical quantities (\ref{rh1}) - (\ref{gibbs}) reduce to the standard Reissner--Nordstr\"om results when $\alpha \to 0$.

\section{Nonlocal Ricci-Tensor and scalar curvature Effects and Classical Black Holes}
\label{sec5}

We will show that in the nonlocal quadratic gravity action (\ref{act}), the propagator kernel $(\Box - \upmu^2)^{-1}$ introduces -- besides the massless graviton -- an additional massive spin-2 excitation which does not correspond to a ghostlike excitation at the classical level.

\subsection{Nonlocal propagator structure and classical stability}
\label{ng}
Expanding the metric $
g_{\mu\nu} = \eta_{\mu\nu} + h_{\mu\nu},$ with $\|h_{\mu\nu}\| \ll 1$,  
and keeping terms quadratic in $h_{\mu\nu}$, the gravitational Lagrangian reads
\begin{equation}
\mathcal L^{(2)} = \frac12 h^{\mu\nu} \mathcal O_{\mu\nu\rho\sigma} h^{\rho\sigma},
\end{equation}
where $\mathcal{O}_{\mu\nu\rho\sigma}$ 
 is the quadratic operator for metric perturbations, 
encoding all kinetic and nonlocal contributions from the action (\ref{act})  (see details in Appendix \ref{a6}). 
Using standard Barnes--Rivers spin projectors, the operator $\mathcal O_{\mu\nu\rho\sigma}$ 
 can be split into spin-0 and spin-2 sectors  \cite{Shapiro:2015uxa,Briscese:2019rii}. 
For the nonlocal quadratic action (\ref{act}), the spin-0 and spin-2 propagators, in the momentum space $\Box \mapsto -p^2$, are respectively given by 
\begin{subequations}
\begin{align}
\Pi_0(p^2) &= \frac{1}{p^2},\\\label{p2p2}
\Pi_2(p^2) &=\frac{p^2 + \upmu^2}{p^2 \left[(1-\alpha)p^2 + \upmu^2\right]}.  
\end{align}
\end{subequations}
The spin-2 propagator (\ref{p2p2}) has exactly two physical spin-2 poles, one massless and one massive, respectively, at
\begin{subequations}
\begin{align}
p^2 &= 0,\label{1st}\\ 
p^2 &= -\frac{\upmu^2}{1-\alpha}.\label{2nd} 
\end{align}
\end{subequations}
The massless pole (\ref{1st}) corresponds to the usual graviton. 
The additional massive spin-2 pole (\ref{2nd}) has residue
\begin{equation}\label{res1}
\textsc{Res.} =
\lim_{p^2 \to -\frac{\upmu^2}{1-\alpha}}
\left(p^2 + \frac{\upmu^2}{1-\alpha}\right)
\frac{p^2 + \upmu^2}{p^2 \left[(1-\alpha)p^2 + \upmu^2\right]}=\frac{\alpha}{1-\alpha},
\end{equation} 
which is positive for $0<\alpha<1$. 
As shown in Appendix~\ref{a6}, the massive spin-2 mode (\ref{2nd}) carries the same sign kinetic term as the massless graviton, 
and therefore it does not correspond to a ghost in the classical regime  \cite{Barvinsky:1994cg,Barvinsky:2003kg}. Therefore, for small values of $\alpha$, the propagator is well behaved, and nonlocal effects yield stable black hole solutions. 

Moreover, it is worth emphasizing that the spin-2 apparent pole 
\begin{equation}
p^2 = -\upmu^2,\label{3rd} 
\end{equation}  whereat the numerator (\ref{p2p2}) cancels, does not produce a separate propagating mode in the spin-2 propagator.  
In fact, at $p^2 \to -\upmu^2$, the propagator (\ref{p2p2}) behaves as  
\begin{equation}
\lim_{p^2\to -\upmu^2}\Pi_2(p^2)
=\lim_{p^2\to -\upmu^2}
\frac{p^2 + \upmu^2}{p^2 \left[(1-\alpha)p^2 + \upmu^2\right]}
=0.
\end{equation}
Hence, it represents a removable singularity rather than a true physical pole, 
and it does not correspond to an independent propagating degree of freedom or a ghost.

\subsection{Phenomenological analysis of classical black holes with nonlocal corrections}

The metric function (\ref{ms1}) represents a leading-order extension of the Reissner--Nordström solution in nonlocal quadratic gravity. The exponential Yukawa-like terms in the metrics (\ref{ms1}, \ref{fm1}) provide a smooth UV modification of the gravitational potential. 
In both the leading-order (\ref{ms1}) and next-to-leading order (\ref{fm1}) metrics, the nonlocal mass scale $\upmu$ controls the exponential Yukawa suppression. A natural phenomenological choice is to relate $\upmu$ to the effective gravitational radius of the hairy black hole, only taking into account the unsuppressed 
$1/r$ term, as
\begin{equation}\label{trm}
\upmu \sim \frac{1}{r_{\textsc{eff}}} = \frac{4\pi}{M},
\end{equation}
Larger [smaller] black holes correspond to smaller [larger] $\upmu$, producing longer-range [shorter-range] nonlocal corrections.  

A solar-mass black hole, $M \sim M_\odot = 1.99 \times 10^{30}\,\mathrm{kg}$ implies that $
r_{\textsc{eff}} \approx 2.95\times 10^{3}\,\mathrm{m}$, 
yielding 
\begin{equation}
\upmu = \upmu_\odot = 6.68\times 10^{-11}\,\mathrm{eV}.\label{sm}
\end{equation}
The mass scale (\ref{sm}) is ultralight from a particle-physics perspective, yet it corresponds to a finite Yukawa screening length $\upmu^{-1}\sim 10^{3}\,\mathrm{m}$. Consequently, nonlocal corrections are relevant only in the near-horizon region of stellar-mass black holes and are exponentially suppressed at larger distances, ensuring consistency with solar-system and galactic-scale tests of gravity. For the supermassive black hole Sagittarius~A$^*$, with mass $M = 4.297\times 10^{6} M_\odot$, one concludes that  
\begin{equation}
\upmu = 1.55 \times 10^{-17}\,\mathrm{eV},
\end{equation}
while for supermassive black holes with masses up to $M \sim 10^{10} M_\odot$, the nonlocal mass scale  is even tinier 
$\upmu \sim 10^{-21}\,\mathrm{eV}$.  
Thus, the nonlocal mass scale $\upmu$ provides a physically meaningful cutoff for Yukawa-type modifications. For a solar-mass black hole, the massive spin-2 pole mass   
\begin{equation}\label{mp1}
m_{\odot, {\scalebox{0.57}{$\textsc{pole}$}}} 
= \frac{6.68}{\sqrt{1-\alpha}} \times 10^{-11}\,\mathrm{eV} 
\end{equation}
is extremely light on particle-physics scales, 
with a Compton wavelength 
\beq\lambda_\textsc{c}=
\frac{\sqrt{1-\alpha}}\upmu = 2.95\,\sqrt{1-\alpha}\times 10^3 \,\,\mathrm{m}.\eeq 
Although the nonlocal quadratic gravity action (\ref{act}), describing a low-energy EFT, is not strictly ghost-free at the quantum level, the massive spin-2 pole (\ref{2nd}) is healthy for $0 < \alpha < 1$, ensuring classical stability. By Eq. (\ref{trm}), the pole mass (\ref{mp1}) is typically far beyond the scale of interest for classical black holes. This allows the leading-order metrics (\ref{ms1}), as well as the next-to-leading order extension  (\ref{fm1}), to remain well-defined classical solutions within a low-energy EFT framework.  

Finally, it is worth noting that in the low-energy EFT governed by the action (\ref{act}), the nonlocal operator admits the geometric series  expansion, as long as the convergence criterion $|\Box| \ll \upmu^2$ is satisfied:
\begin{equation}
(\Box-\upmu^2)^{-1}
=
- \frac{1}{\upmu^2} \sum_{n=0}^{\infty} \left(\frac{\Box}{\upmu^2}\right)^n.\label{large-mu-expansion}
\end{equation}
As a consequence, the limit 
\beq\label{large-mu-limit}
\displaystyle
\lim_{\upmu\to\infty}(\Box-\upmu^2)^{-1}
=
-\frac{1}{\upmu^2}\eeq makes the heavy degree of freedom infinitely massive, so that it decouples from the low-energy
dynamics. The nonlocal part of the action \eqref{act} thus reduces to the local Ricci squared invariant:
\beq\label{nl2}
\lim_{\upmu\to\infty}S_{\scalebox{0.57}{$\textsc{NL}$}}
=-\frac{\alpha}{\upmu^2}\int d^4x\,\sqrt{-g}
\left(R_{\mu\nu}R^{\mu\nu}-\frac14R^2\right),\eeq
while the inverse operator limit \eqref{large-mu-limit} shrinks the interaction to zero distance. It shows that nonlocal gravity smoothly reduces to local quadratic gravity when $\upmu\to\infty$, with all
corrections suppressed by powers of $\upmu^{-2}$ in this decoupling limit.

\section{Conclusions}
\label{sec6}
In this work, we constructed hairy black hole solutions in nonlocal quadratic gravity. This framework provides a natural setting to explore the interplay between black hole physics and nonlocal effects. The metric receives leading-order corrections (\ref{ms1}) and next-to-leading nonlocal contribution (\ref{fm1}), which together systematically extend the Reissner--Nordström geometry with Yukawa-type corrections. 
We have demonstrated that nonlocal corrections implement a Yukawa screening effect. In fact, curvature contributions of the hairy black hole are effectively smeared over a short range due to nonlocal Yukawa-type interactions. The hairy black hole behaves as if its mass is slightly less concentrated near the horizon, yielding the event horizon to shift inward.  This naturally leads to a modified Hawking temperature and also reduces the Bekenstein--Hawking entropy. Nonlocal corrections slightly stabilize small charged black holes by decreasing the magnitude of the negative specific heat capacity, while renormalizing the chemical potential. The nonlocal-corrected Helmholz and Gibbs energies support the absence of first-order phase transitions. 

The nonlocal propagator analysis and classical stability have also been  addressed. The spin-2 propagator contains, besides a massless graviton, a massive pole (\ref{2nd}) corresponding to a healthy mode that is not a ghost at the classical level. Phenomenologically, the hairy black hole metrics (\ref{ms1}) and (\ref{fm1}) are highly versatile, with variations in mass, charge, and hair driven by nonlocal effects. Yukawa-type corrections make them useful for studies of nonlocal gravity and for exploring potential observational signatures. 
\medbreak
{\textbf{Acknowledgements}}: 
RdR is supported by The S\~ao Paulo Research Foundation (FAPESP) 
(Grants No. 2021/01089-1 and No. 2024/05676-7) and the National Council for Scientific and Technological Development (CNPq) (Grants No. 303742/2023-2 and No. 401567/2023-0). 
\appendix

\section{Details of the spin-2 propagator, poles, and residue analysis}\label{a6}

We showed in Subsec. \ref{ng} that the spin-2 propagator has a massless pole (\ref{1st}) at $p^2 = 0$,  corresponding to the usual graviton, and a massive spin-2 pole (\ref{2nd}) at $p^2 = -\upmu^2/(1-\alpha)$, whose overall sign of the residue (\ref{res1}) is consistent with the kinetic term in the classical action. 
For the physically relevant regime $0 < \alpha < 1$, the relative sign of the residue with respect to the massless graviton is fixed by the Einstein--Hilbert term in the action (\ref{act}). In fact, starting from the nonlocal part (\ref{nl1}) of the action (\ref{act}), 
one expands the metric around flat spacetime, 
$g_{\mu\nu}=\eta_{\mu\nu}+h_{\mu\nu}$, with $\|h_{\mu\nu}\| \ll1$. 
At quadratic order, the action can be written in momentum space as
\begin{equation}\label{qa22}
S^{(2)}=\frac12\int d^4p\;
h^{\mu\nu}(-p)\,
\mathcal O_{\mu\nu\rho\sigma}(p)
h^{\rho\sigma}(p),
\end{equation}
where the operator $\mathcal O_{\mu\nu\rho\sigma}$ is decomposed using the Barnes--Rivers spin projectors \cite{Barvinsky:1994cg,Barvinsky:2003kg}. 
Due to the specific combination of nonlocal terms in the nonlocal part  (\ref{nl1}) of the action (\ref{act}),  
the spin-0 sector cancels identically, and only the spin-2 sector is modified. 
For spin-2 modes, the action (\ref{qa22}) takes the form
\begin{equation}\label{zko}
S^{(2)}_{\scalebox{0.57}{$\textsc{spin-2}$}} 
= \frac12 \int d^4p \; 
h^{(2)}_{\mu\nu}(-p)  
P^{(2)\mu\nu\rho\sigma}  
\frac{p^2 \big[ (1-\alpha)p^2 + \upmu^2 \big]}{p^2 + \upmu^2} 
h^{(2)}_{\rho\sigma}(p),
\end{equation}
where $P^{(2)\mu\nu\rho\sigma}$ is the spin-2 Barnes--Rivers projector
\begin{equation}
P^{(2)\mu\nu\rho\sigma}(p)
= \frac12 \Big( \mathcal{P}^{\mu\rho} \mathcal{P}^{\nu\sigma} + \mathcal{P}^{\mu\sigma} \mathcal{P}^{\nu\rho} \Big) 
- \frac{1}{3} \mathcal{P}^{\mu\nu} \mathcal{P}^{\rho\sigma},
\end{equation} and the transverse projector reads
$
\displaystyle\mathcal{P}_{\mu\nu}(p) = \eta_{\mu\nu} - \frac{p_\mu p_\nu}{p^2}.
$
The zeroes of the kinetic operator (\ref{zko}) occur at the poles (\ref{1st}, \ref{2nd}),
corresponding to the massless graviton and an additional massive spin-2 mode, respectively.

Near the massless pole (\ref{1st}), the quadratic action reduces to
\begin{equation}
S^{(2)}_{\scalebox{0.57}{$\textsc{graviton}$}}
= \frac12 \int d^4p \;
h^{(2)}_{\mu\nu}(-p)  P^{(2)\mu\nu\rho\sigma}  p^2  h^{(2)}_{\rho\sigma}(p),
\end{equation}
which fixes the positive normalization of the graviton kinetic term. 
On the other hand, expanding the action near the massive pole (\ref{2nd}), one finds
\begin{equation}\label{pfa}
S^{(2)}_{\scalebox{0.57}{$\textsc{massive}$}}
= \frac{1-\alpha}{2\alpha} \int d^4p \;
h^{(2)}_{\mu\nu}(-p)  
P^{(2)\mu\nu\rho\sigma}  
\left( p^2 + \frac{\upmu^2}{1-\alpha} \right) 
h^{(2)}_{\rho\sigma}(p).
\end{equation}
For $0<\alpha<1$, the prefactor $(1-\alpha)/2\alpha$ in the action (\ref{pfa}) is strictly positive. 
Therefore, the massive spin-2 mode carries the same sign kinetic term as the massless graviton, and does not correspond to a ghost.  
Thus,  nonlocal modifications yield a propagator that is well-defined and stable at classical scales within the EFT regime.

\renewcommand{\bibfont}{\footnotesize}
\bibliographystyle{apsrev}
\bibliography{gdq}

\end{document}